\documentclass[preprint,aps,amsmath,pre,showpacs]{revtex4-1}
\usepackage{graphicx}
\newcommand{\dps}{\displaystyle}

\begin{document}

\title{Binary dynamics on star networks under external perturbations}

\author{Carolina A. Moreira, David M. Schneider and Marcus A.M. de Aguiar}
\email{corresponding author:aguiar@ifi.unicamp.br}

\affiliation{Instituto de F\'{\i}sica `Gleb Wataghin',
Universidade Estadual de Campinas, Unicamp\\ 13083-970, Campinas,
SP, Brazil}

\begin{abstract}

We study a binary dynamical process that is a representation of the voter model
with two candidates and opinion makers. The voters are represented by nodes of
a network of social contacts with internal states labeled 0 or 1 and nodes
that are connected can influence each other. The network is also perturbed by
opinion makers, a set of external nodes whose states are frozen in 0 or 1 and
that can influence all nodes of the network. The quantity of interest is the
probability of finding $m$ nodes in state 1 at time $t$. Here we study this
process on star networks, which are simple representations of hubs found in
complex systems, and compare the results with those obtained for networks that
are fully connected. In both cases a transition from disordered to ordered
equilibrium states is observed as the number of external nodes becomes small.
For fully connected networks the probability distribution becomes uniform at the
critical point. For star networks, on the other hand, we show that the
equilibrium distribution splits in two peaks, reflecting the two possible states
of the central node. We obtain approximate analytical solutions for the
equilibrium distribution that clarify the role of the central node in the
process. We show that the network topology also affects the time scale of
oscillations in single realizations of the dynamics, which are much faster for
the star network. Finally, extending the analysis to two stars we compare our
results with simulations in simple scale-free networks.

\end{abstract}

\pacs{89.75.-k,05.50.+q,05.45.Xt}
%89.75.-k    Complex systems
%05.50.+q    Lattice theory and statistics (Ising, Potts, etc.)
%05.45.Xt    Synchronization; coupled oscillators

%keywords    networks; finite systems; ising model; external perturbations

\maketitle

\section{Introduction}
\label{intro}

Network science has provided a large body of theoretical tools to investigate
complex systems, from physics to social sciences and biology
\cite{bararev,hwa06,murray,newman10,sporns11,gross2011}. Much work has been
devoted to the study of networks topological properties 
\cite{bararev,murray,newman10,baryam2004,alb,cohen,buldyrev10} and dynamical
processes on networks have been shown to depend sensitively on the network structure 
\cite{barrat08,ves01,pecora02,motter03,pac04,laguna,guime,aguiar05,rodrigues13,
rodrigues14}. More recently, the response of networks to external perturbations
has also been investigated
\cite{baryam2004,hill2010,xing,lee,marketcrises11,chinellato2015}.

Most of the networks found in nature are scale-free, characterized by a
power-law degree distribution and by the presence of nodes whose degree greatly
exceeds the average \cite{bararev}. These special nodes, referred to as network
hubs, are crucial for the structural integrity of many real-world systems
\cite{quax2013}, allowing for a fault tolerance behavior against random
failures \cite{alb}. Nevertheless, if the hubs are removed from
the network by an intentional attack, the network might fragment into a set of
isolated graphs. Thus, the presence of hubs represents at the same time the
robustness and the  `Achilles heel' of scale-free networks. This property has
been extensively studied by means of percolation theory \cite{cohen,
cohen2001,callaway2000}. In addition, network hubs can be detected and studied
using numerous different graph measures, most of which express aspects of node
centrality \cite{Martijn}. 

In this paper we study the two-states Voter Model subjected to external
perturbations in star networks and compare the results with those obtained for
fully connected systems. The star and fully connected topologies model two
extreme scenarios, corresponding to the presence of a single network hub and the
total absence of preferentially connected nodes, respectively. The perturbations
represent opinion makers, who have already decided who to vote for and whose
influence extents over the entire population. They are modeled by a set of
external nodes whose states are fixed and that connect to all nodes of the
network. The system exhibits a phase transition from disordered to ordered
states as the external perturbation is decreased and can be characterized by the
equilibrium probability distribution of finding $m$ nodes in a given state. We
show that the shape of this distribution is very similar for star and fully
connected networks away from the phase transition, but it shows a fingerprint of
the network topology close to the critical point. For fully connected networks
the probability distribution is uniform at the critical point, whereas for  star
networks it splits in two peaks, reflecting the two possible states of the
central node. For single realizations of the dynamics and weak perturbations the
state of the network oscillates according to the equilibrium distribution. We
show that the time scale of these oscillations is sensitive to the network
topology, being much faster for the star network. We derive approximate
analytical solutions for the star network and extend the results for multiple
stars, which can be used as a simplified model for a scale-free network.

In the next two sections we describe the Voter Model with opinion makers and the
implementation of the dynamics in a general network. Exact results for the fully
connected network are reviewed in section \ref{fullynet} and in section
\ref{starnet} we obtain the master equations for star networks and show results
from numerical simulations. We also generalize our calculations to star
networks whose center contains a group of fully connected nodes, and construct
an approximate solution for the joint effect of two network hubs, which is
further compared with the outcome of a scale-free network. Our conclusions are
presented in section \ref{conclusions}.

\section{The voter model}\label{voter}

The voter model consists of a set of individuals trying to decide in which of
two candidates to vote for \cite{liggett1985,acemoglu2013}. Their opinion can be
influenced by their friends, represented by a network of social contacts, and by
opinion makers, such as journalists or politicians, whose power of persuasion
toward one of the candidates extents over the entire population. The opinion
makers are modeled by external 'frozen' nodes whose states never change
and that reach all voters equally, acting as perturbations to the 
network dynamics (Fig. \ref{fig1}). 

The intention of a voter is quantified by its state being 0 or 1 and the number
of opinion makers for candidates 0 and 1 are $N_0$ and $N_1$ respectively. At
each time step a voter is selected at random and its state is updated: the voter
can retain its opinion or adopt the opinion of one of its connected neighbors,
which can be a friend or an opinion maker. In the absence of opinion makers the
population eventually reaches a consensus and the network stabilizes with all
nodes 0 or all nodes 1, which are the only absorbing configurations. As long as opinion
makers are present for both candidates the network never stabilizes, but it does
reach a statistical equilibrium where the probability that candidate 1 has a
given number of votes becomes independent of the time.

This dynamical process can model other interesting systems besides as an
election with two candidates \cite{vilone,redner}, such as a population of
sexually reproducing (haploid) organisms \cite{moran,aguiar2011} and herding
behavior in social systems \cite{kirman,marketcrises11}. It is also similar to
the Glauber dynamics of the Ising model \cite{glauber,chinellato2015} where
$N_0+N_1$ is analogous to the temperature and $N_0-N_1$ to an external magnetic
field.

If the number of opinion makers is zero the average time to reach consensus can
be analytically calculated in terms of the moments of the network degree
distribution \cite{sood2008,vazquez2008}. However, the presence of external
perturbations complicates the dynamics and solutions have been obtained only for
simple networks and specific distribution of frozen nodes. In particular, the
voter model without opinion makers was studied in regular lattices  where one
individual in the population has fixed opinion (a zealot) \cite{mobilia2003}.
Analytic solutions were also obtained for the equilibrium distribution in fully
connected networks with arbitrary number of opinion makers in the limit where
the number of voters goes to infinity \cite{kirman,redner2007}. The full
dynamical problem with finite number of voters was finally solved in
\cite{chinellato2015} where it was shown that the solution was also a good
approximation for networks of different topologies, as long as the number of
opinion makers $N_0$ and $N_1$ were rescaled according to the average degree of
the network (see also \cite{alfarano}). The numbers $N_0$ and $N_1$ were also
analytically extended to real numbers smaller than 1, representing weak coupling
between the voters and the opinion makers. It was shown (see also
\cite{kirman,redner2007}) that a phase transition exists between ordered states,
where most voters have the same opinion, to a disordered state, where
approximately half the votes go to each candidate, as $N_0$ and $N_1$ go from
very small to very large numbers. The transition occurs exactly at $N_0=N_1=1$
for fully connected networks of any size. Here we study this phase transition in
the star network.

% % % % % % % % % % % % % % % % % % % % % % % % % % % % % % %% % % % %
% % % % % % % % % % % % % % % % % % % % % % % % % % % % % % % % % % %
\section{Network dynamics} \label{model}

\begin{figure}
   \includegraphics[clip=true,width=10cm]{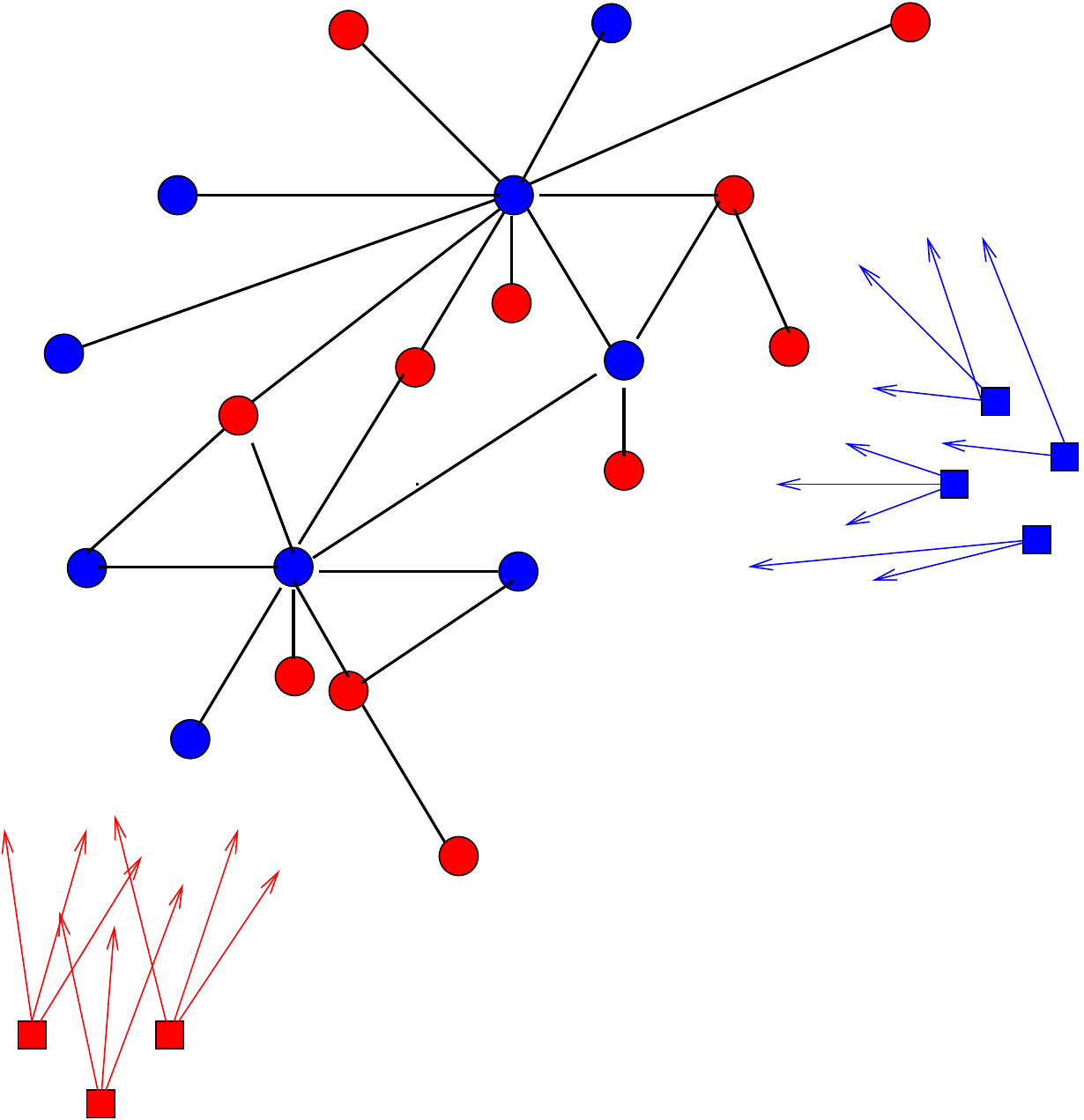}
   \caption{(color online) Representation of the voter model on a network. The 
   different colors indicate the internal states of the voters, which can be undecided (circles), 
   and opinion makers (squares). Opinion makers 
   can affect all voters of the network but undecided voters can only influence 
   their connected neighbors.}
  \label{fig1}
\end{figure}

Consider a network with $N$ nodes specified by the adjacency matrix $A$, defined
by $A_{ij}=1$ if nodes $i$ and $j$ are connected and $A_{ij}=0$ otherwise. For
our purpose, $A_{ii}=0$ (nodes do not connect to themselves), and for any pair
of nodes it is possible to construct a path connecting them.  Each node has an
internal state which can take only the values $0$ or $1$. The nodes are also
connected to $N_0$ external nodes whose states are fixed at 0 and to $N_1$ nodes
whose states are fixed at 1, as illustrated in Fig.\ref{fig1}. In order to
distinguish between the two kinds of nodes, we call the $N_0+N_1$ external nodes
{\it fixed} and the $N$ nodes of the network, whose states are variable, {\it
free}. Following \cite{chinellato2015} we shall treat $N_0$ and $N_1$ as
real numbers, representing weighted coupling between the opinion
makers and the voters. 

The free nodes can change their internal state according to the following
dynamical rule: at each time step a free node is selected at random and, with
probability $p$ its state remains the same; with probability $1-p$ the node
copies the state of one of its connected neighbors, free or fixed, also chosen
at random.

Let
\begin{equation}
 x=\{x_1,x_2,\dots,x_{k-1},x_k,x_{k+1},\dots,x_N\}
\end{equation}
denote a {\it microscopic} state of the network with $x_i=0$ or $x_i=1$ representing
the state of node $i$. There is a total of $2^N$ possible microscopic states and
we call $P_t(x)$ the probability of finding the network in the state $x$ at time
$t$. Since a single free node can change state per time step, it is useful to
define the auxiliary state $x^k$ which is identical to $x$ at every node except
at node $k$, whose state is the opposite of $x_k$, i.e., $x^k_k = 1 - x_k$.
Explicitly,
\begin{equation}
 x^k=\{x_1,x_2,\dots,x_{k-1},1-x_k,x_{k+1},\dots,x_N\}.
\end{equation}

With these definitions, the evolution equation for the probabilities can be
written as
\begin{equation}
\begin{array}{ll}
P_{t+1}(x) = p P_t(x) & + (1-p)\frac{1}{N} P_t(x) \sum_{i=1}^N T(x_i \rightarrow
x_i) \\
&+ (1-p) \sum_{i=1}^N \frac{1}{N} P_t(x^i) T(x^i_i \rightarrow x_i).
\end{array}
\end{equation}
The first two terms take into account the probability that the network is
already in state $x$ and the selected node (i) does not change its state or (ii)
copies the state of a neighbor which is identical to its own state. The last
term is the probability that the network is in a state differing from $x$ by a
single node, which is selected and copies the state of neighbor opposite to its
own.

According to the dynamical rules, the transition probabilities can be written as
\begin{equation}
T(x_i \rightarrow x_i) = \frac{1}{k_i+N_0+N_1}\left[\sum_{j=1}^N A_{ij}
|1-x_i-x_j| + x_i N_1 + (1-x_i) N_0 \right]
\end{equation}
and
\begin{equation}
T(x^i_i \rightarrow x_i) = \frac{1}{k_i+N_0+N_1}\left[\sum_{j=1}^N A_{ij}
|x^i_i-x_j| + (1-x^i_i) N_1 + x^i_i N_0 \right]
\end{equation}
where $k_i$ is the degree of the node $i$. Using the fact that $x^i_i=1-x_i$ we find
that the two transition probabilities are identical and obtain 
\begin{equation}
\begin{array}{ll}
P_{t+1}(x) &= p P_t(x)  + \\
& \dps{\frac{(1-p)}{N} \sum_{i=1}^N \frac{[P_t(x)+P_t(x^i)]}{k_i+N_0+N_1}
\left[\sum_{j=1}^N A_{ij} |1-x_i-x_j| + x_i N_1 + (1-x_i) N_0\right]}.
\end{array}
\label{geneq}
\end{equation}
%

% % % % % % % % % % % % % % % % % % % % % % % % % % % % % % % % % % %
% % % % % % % % % % % % % % % % % % % % % % % % % % % % % % % % % % %
\section{Fully connected networks}
\label{fullynet}

For networks that are fully connected the nodes are indistinguishable and the
state of the network is fully specified by the number $m$ of nodes with internal
state 1 \cite{aguiar05,chinellato2015}. Each of these {\it macroscopic} states
corresponds to a set of $N!/[(N-m)! m!]$ degenerated microscopic network states.
Because there are only $N+1$ macroscopic states equations (\ref{geneq}) are
greatly simplified. The equilibrium probability $\rho_{FC}(m)$ of finding the
network with $m$ nodes in state 1 is given by the Beta-Binomial distribution
\cite{chinellato2015}
\begin{equation}
\rho_{FC}(m) =  A(N,N_0,N_1) ~ \frac{\Gamma(N_1+m)~\Gamma(N + N_0 - m)}
{\Gamma(N-m+1)~\Gamma(m+1)},
\label{probn}
\end{equation}
where
\begin{equation}
A(N,N_0,N_1) = \frac{\Gamma(N+1)~\Gamma(N_0+N_1)}{\Gamma(N + N_0 + N_1)
~\Gamma(N_1)~\Gamma(N_0)}.
\end{equation}
This expression can also be written in term of $x_m = m/N$. In the limit
$N\rightarrow \infty$, $x_m$ becomes a continuous variable $0 \leq x \leq 1$
and $\rho_{FC}$ converges to the Beta distribution \cite{kirman}
\begin{equation}
\rho_{FC}(x) =  \frac{\Gamma(N_0+N_1)}{\Gamma(N_0) \Gamma(N_1)} x^{N_0-1}
(1-x)^{N_1-1}.
\label{probbetax}
\end{equation}
The interesting feature of the solution expressed by Eq.(\ref{probn}) is that
for $N_0=N_1=1$ it gives $\rho_{FC}(m)=1/(N+1)$, meaning that all states are
equally likely, as illustrated in Fig.\ref{fig2}.

For networks of different topologies the effect of the fixed nodes is amplified.
The probability that a free node copies a fixed node is
$P_i=(N_{0}+N_{1})/(N_{0}+N_{1}+k_i)$, where $k_i$ is the degree of the node.
For fully connected networks $k_i=N-1$ and $P_{FC} \equiv
(N_{0}+N_{1})/(N_{0}+N_{1}+N-1)$. For general networks an average value $P_{av}$
can be calculated by replacing $k_i$ by the average degree. Effective numbers of
fixed nodes $N_{0ef}$ and $N_{1ef}$ can be then defined as the values of $N_0$
and $N_1$ in $P_{FC}$ for which $P_{av} \equiv P_{FC}$. This leads to 
\begin{equation}
N_{0ef} = f N_0, \qquad \qquad N_{1ef} = f N_1,
\label{eq:rescaling}
\end{equation}
where $f=(N-1)/k_{av}$. In \cite{chinellato2015} it was shown that
Eq.(\ref{probn}) with the above rescaling of fixed nodes fits very well the
probability distribution for a variety of topologies. The formula was tested for
relatively small networks of the types random, 2-D regular lattice,
Barabasi-Albert scale-free and small world. Similar results were obtained in the
context of herding behavior of economic agents \cite{alfarano,marketcrises11}.

% % % % % % % % % % % % % % % % % % % % % % % % % % % % % % % % % % %
% % % % % % % % % % % % % % % % % % % % % % % % % % % % % % % % % % %
\section{Star networks}
\label{starnet}

% % % % % % % % % % % % % % % % % % % % % % % % % % % % % % % % % % %
\subsection{Master equation}

For a star network it is convenient to set the total number of nodes to $N+1$.
Node 1 is the central node and it is connected to all peripheral $N$ nodes. The
peripheral nodes, on the other hand, are only connected to the central node. The
peripheral nodes are indistinguishable from each other and, similar to the fully
connected network, there are only $2(N+1)$ macroscopic states, characterized by
having $m$ peripheral nodes in state 1 ($N+1$ possibilities) and the central
node in state 1 or 0.

The evolution equation for the macroscopic states can be obtained from equations
(\ref{geneq}) if we define  $r_1(m,t)$ and $r_0(m,t)$ as the probabilities of
having $m$ peripheral nodes in state 1 at time $t$ with the central node in
state 1 and 0 respectively. We obtain
\begin{align}
\begin{split}
r_1(m, t+1) &= r_1(m,t) \left\{ p + \frac{(1-p)}{(N+1)} \left[ \frac{m (N_1 + 1) + 
(N-m) N_0 }{(1+N_1+N_0)}  + \frac{(m+N_1)}{(N + N_0 + N_1)} \right]  \right\} \\
& + r_1(m+1, t) \,(1-p)\frac{(m+1) N_0}{(N+1)(1+N_0+N_1)}  \\
& + r_1(m-1,t)  \,(1-p)\frac{(N-m+1) (N_1 + 1)}{(N+1)(1+N_0+N_1)}   \\
& + r_0(m,t) \,(1-p)\frac{(m+N_1)}{(N+1)(N+N_0+N_1)},
\label{eqr}
\end{split}
\end{align}
and
\begin{align}
\begin{split}
r_0(m, t+1) &= r_0(m,t) \left\{ p + \frac{(1-p)}{(N+1)} \left[ \frac{m N_1 +
(N-m) (N_0+1) }{(1+N_1+N_0)}  + \frac{(N-m+N_0)}{(N + N_0 + N_1)} \right]
\right\} \\
& + r_0(m+1, t)\,(1-p) \frac{(m+1) (N_0+1)}{(N+1)(1+N_0+N_1)}  \\
& + r_0(m-1,t)  \, (1-p)\frac{(N-m+1) N_1}{(N+1)(1+N_0+N_1)}   \\
& + r_1(m,t) \, (1-p)\frac{(N-m+N_0)}{(N+1)(N+N_0+N_1)}.
\end{split}
\label{eqs}
\end{align}

The first two terms in these equations take into account the probability that the 
network is in the state $x$ at time $t$, and to select a node that (i) does not
change its state or (ii) copies the state of a neighbor in its own state.
The last three terms represent the probability that the network is in a state
differing from $x$ by a single node at time $t$, to select this node, and to
copy the state of a neighbor in the opposite state.

\begin{figure}
   \includegraphics[clip=true,width=8cm]{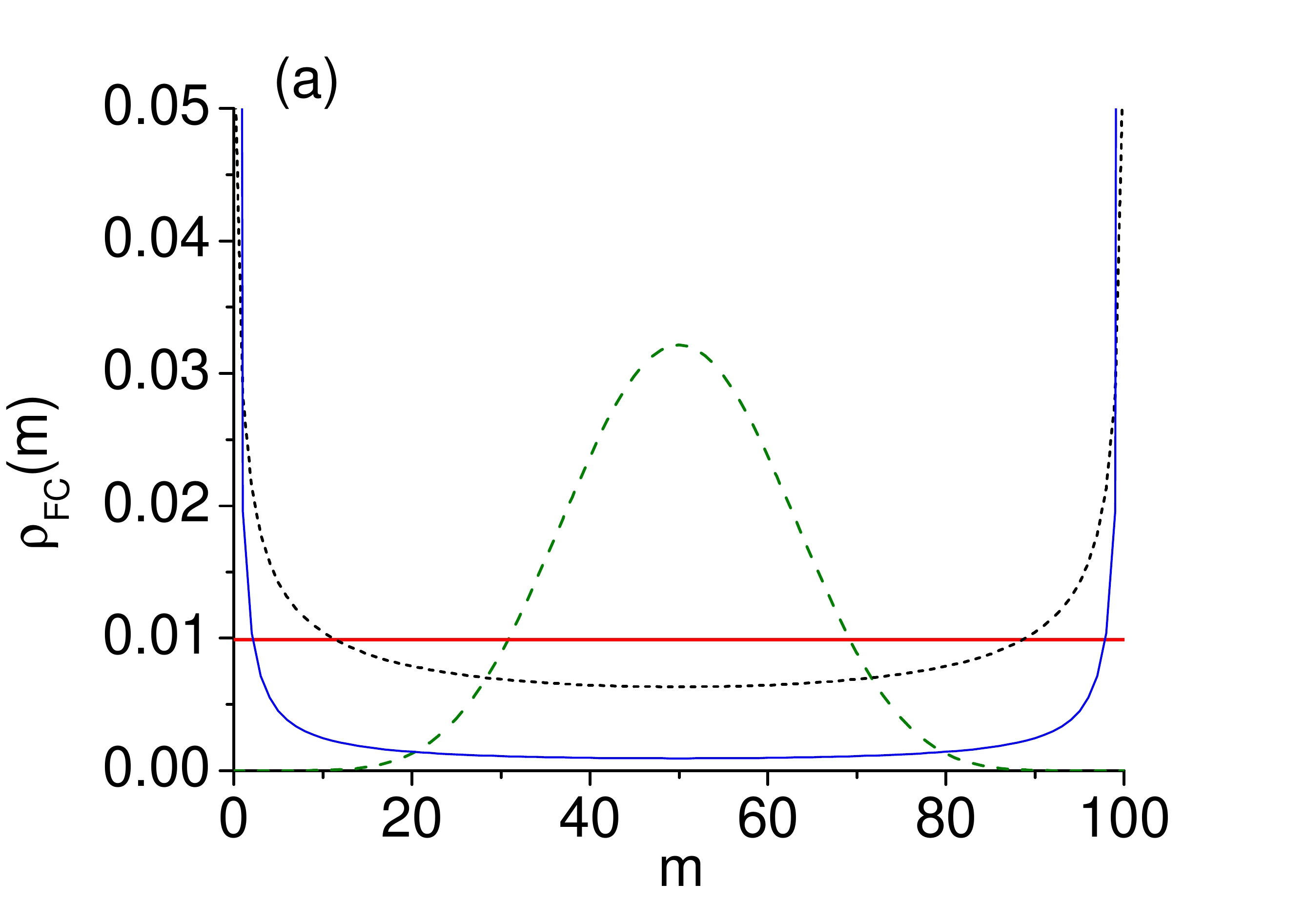}
   \includegraphics[clip=true,width=8cm]{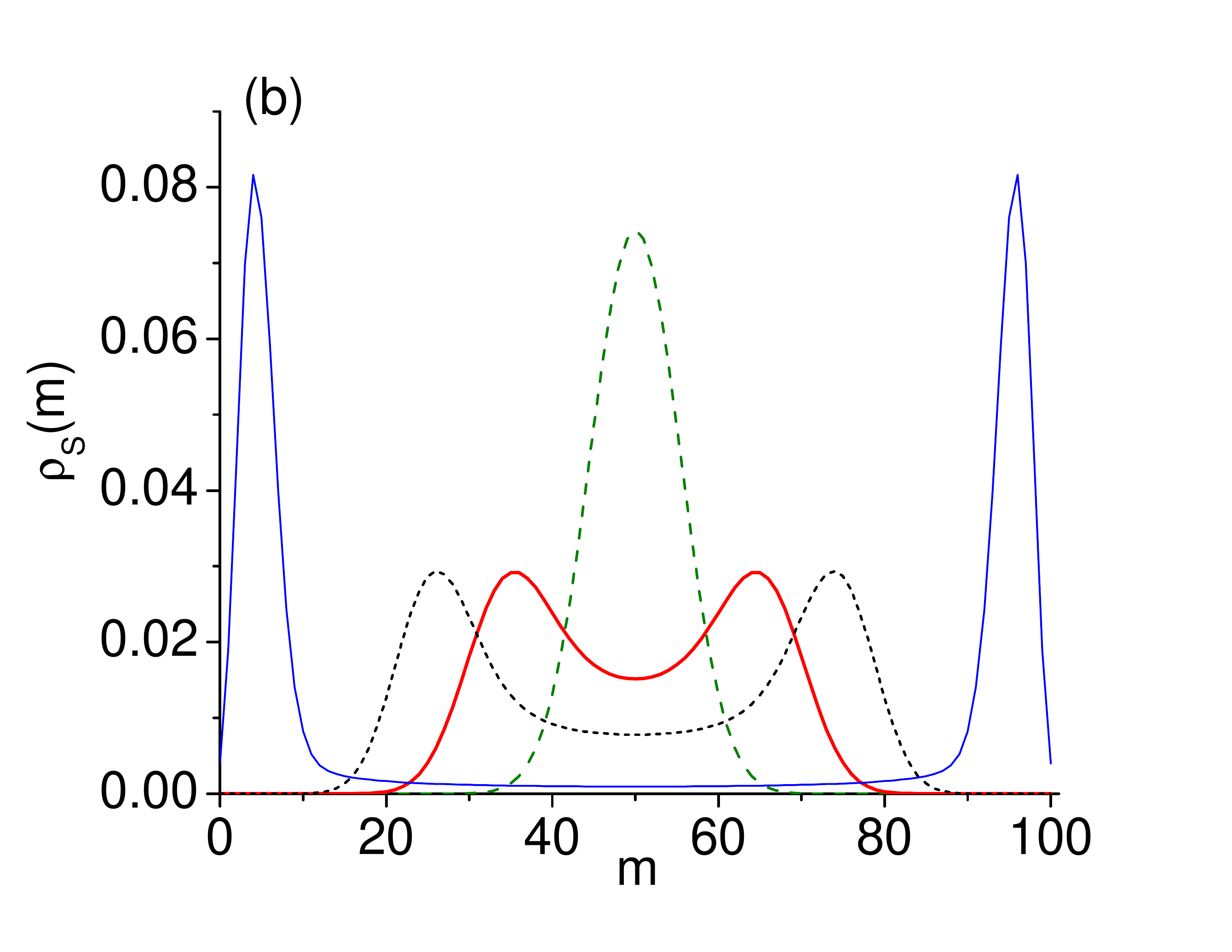}
   \caption{(color online) Equilibrium probability distribution for networks
  with $N=100$ nodes and different values of $N_p=N_0=N_1$: $N_p=10$ (green
  dashed), $N_p=1$ (thick red), $N_p=0.5$ (black dotted), $N_p=0.05$ (blue). 
  (a) fully connected network; (b) star network. }
  \label{fig2}
\end{figure}

The probability of having $m$ nodes in state 1 in the star network is,
therefore,
\begin{equation}
\rho_S(m,t)=r_1(m-1,t)+r_0(m,t).
\label{eqpmstar}
\end{equation}
The results provided by equations (\ref{eqr}) and (\ref{eqs}) agree perfectly
well with numerical simulations. A comparison with the fully connected network
is shown in Fig.\ref{fig2}. The main feature of these results is the different
way in which the transition between ordered and disordered states occurs:
instead of the meltdown of the Gaussian distribution observed for fully
connected networks, the Gaussian state splits in two peaks that move toward the
boundaries $m=0$ and $m=N$ as $N_0$ and $N_1$ are decreased.

% % % % % % % % % % % % % % % % % % % % % % % % % % % % % % % % % % %
\subsection{Approximate solutions}

The main difficulty in solving equations (\ref{eqr}) and (\ref{eqs}) is that
they are coupled through the central node. Although we have not found exact
solutions, a simple enough approximation can be readily obtained if the central
node is momentarily considered to be fixed. If the central node is fixed in
state 1, any peripheral node sees $N_0$ fixed nodes in state 0 and $N_1+1$ nodes
fixed in state 1. The problem reduces to that of $N$ independent nodes. The
asymptotic probability that a peripheral node is in state 1 is 

\begin{figure}
   \includegraphics[clip=true,width=8cm]{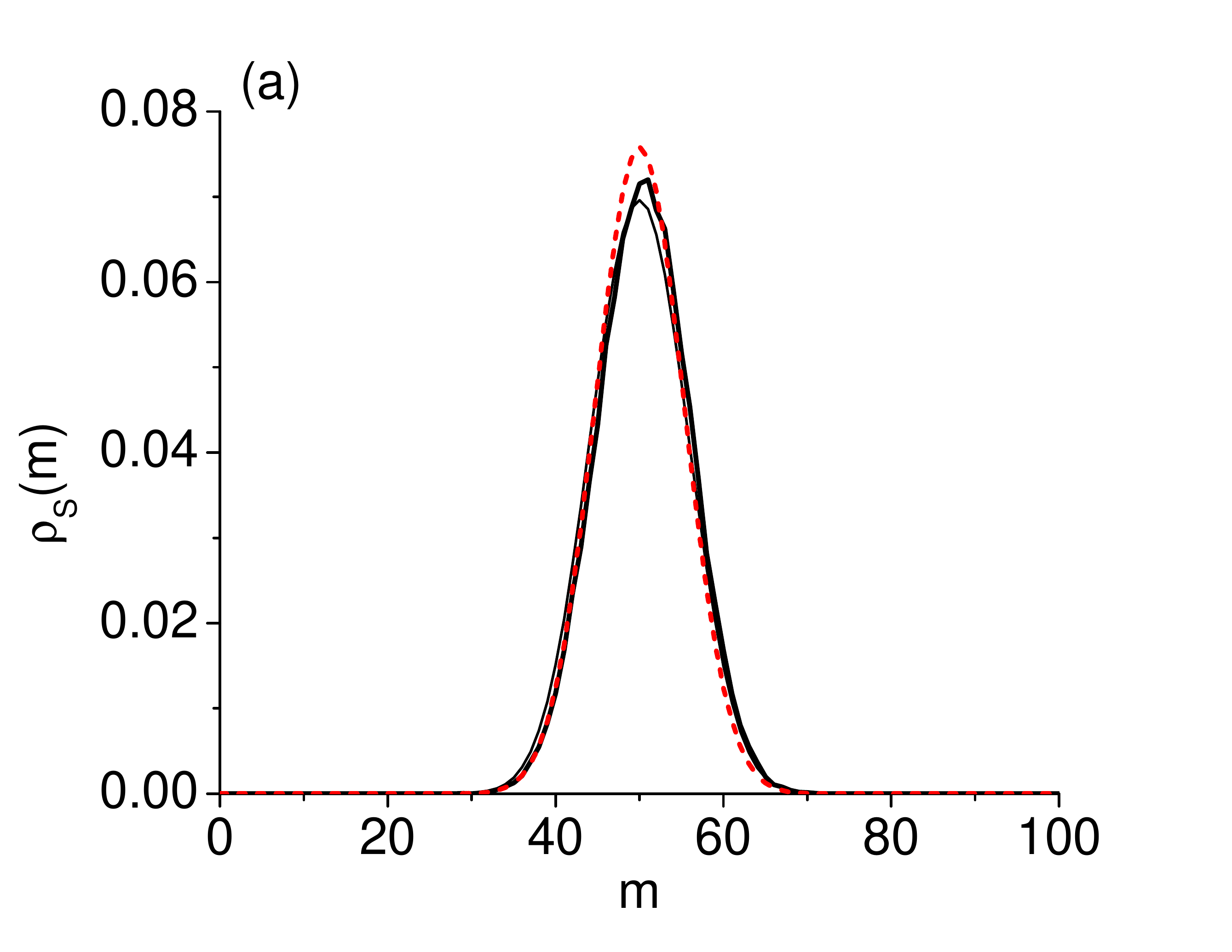}
   \includegraphics[clip=true,width=8cm]{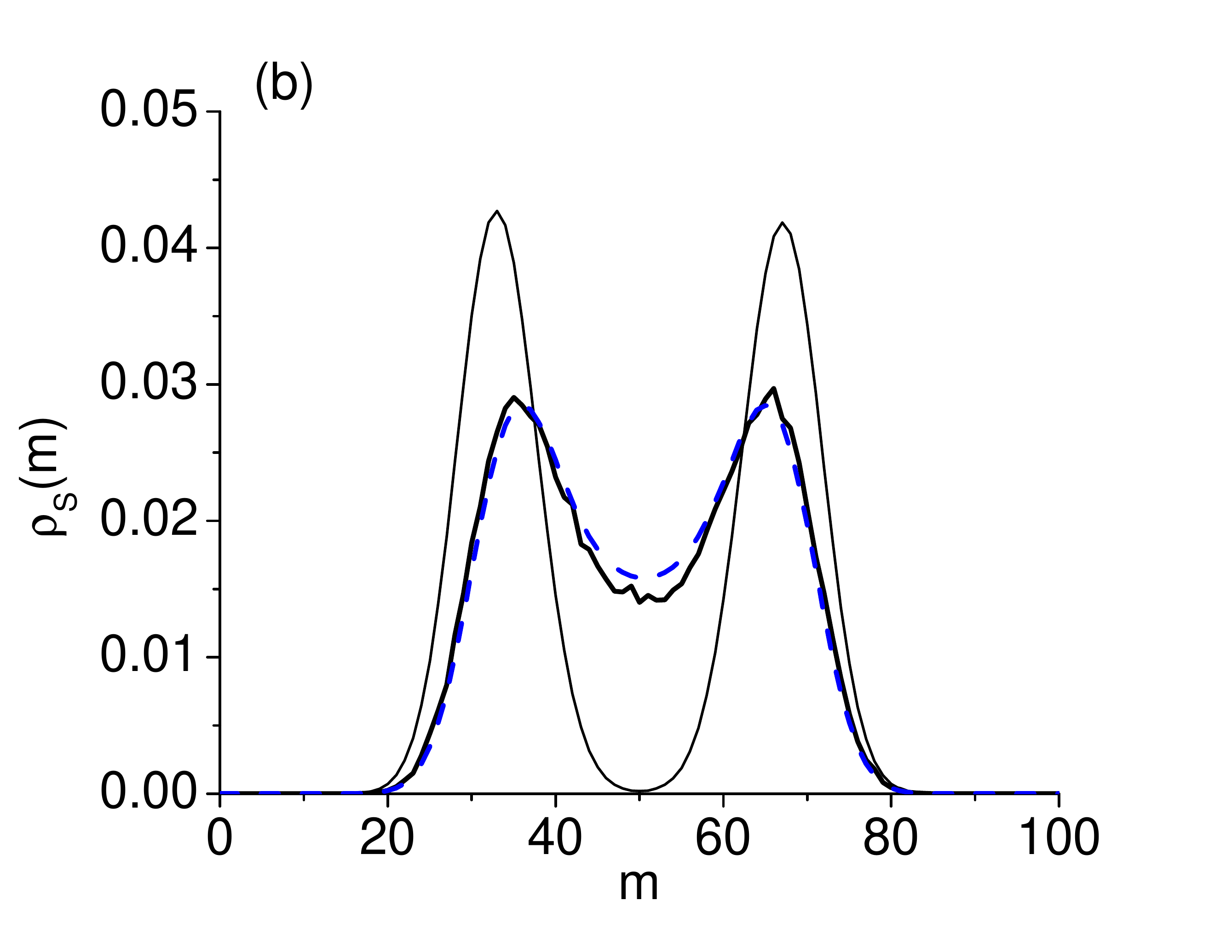}
   \includegraphics[clip=true,width=8cm]{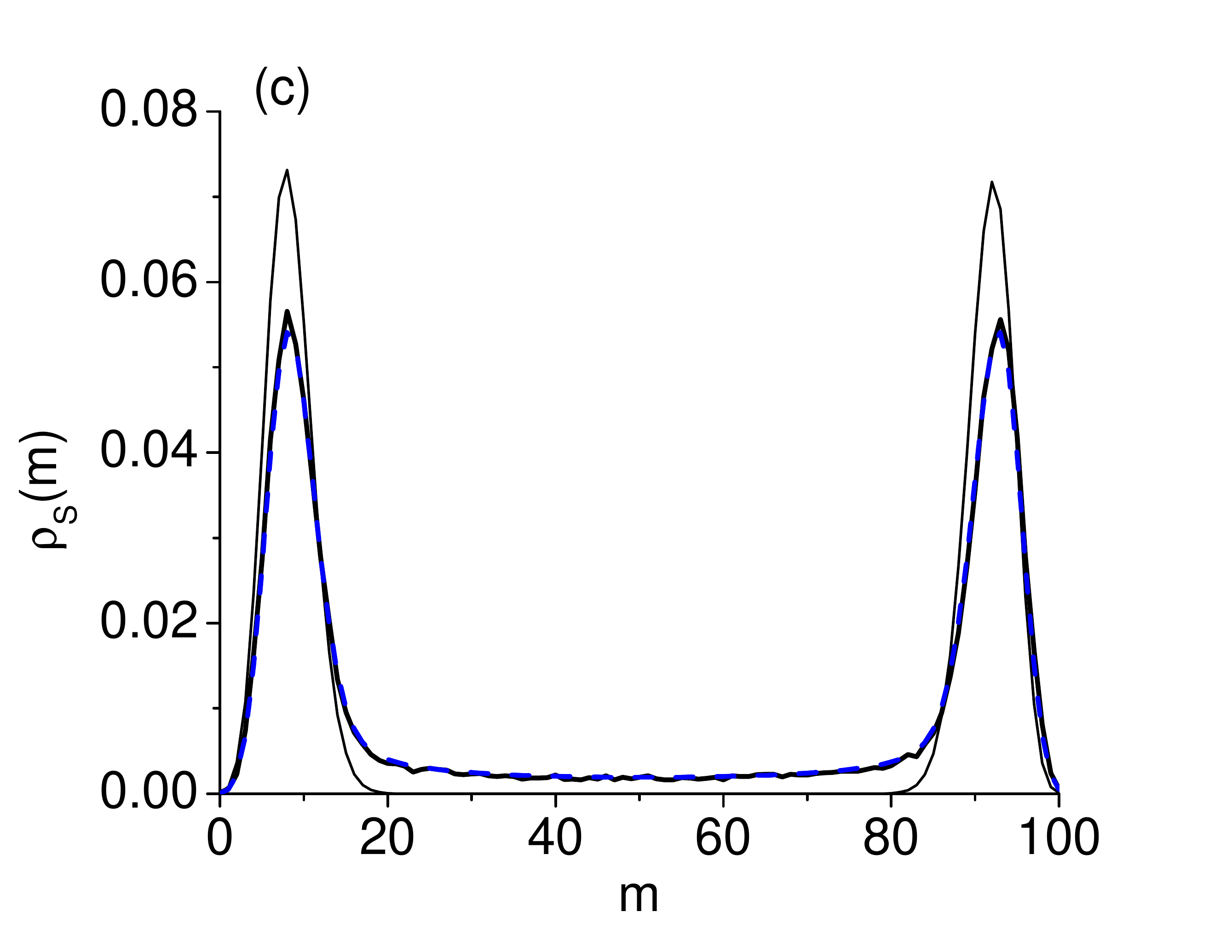}
   \includegraphics[clip=true,width=8cm]{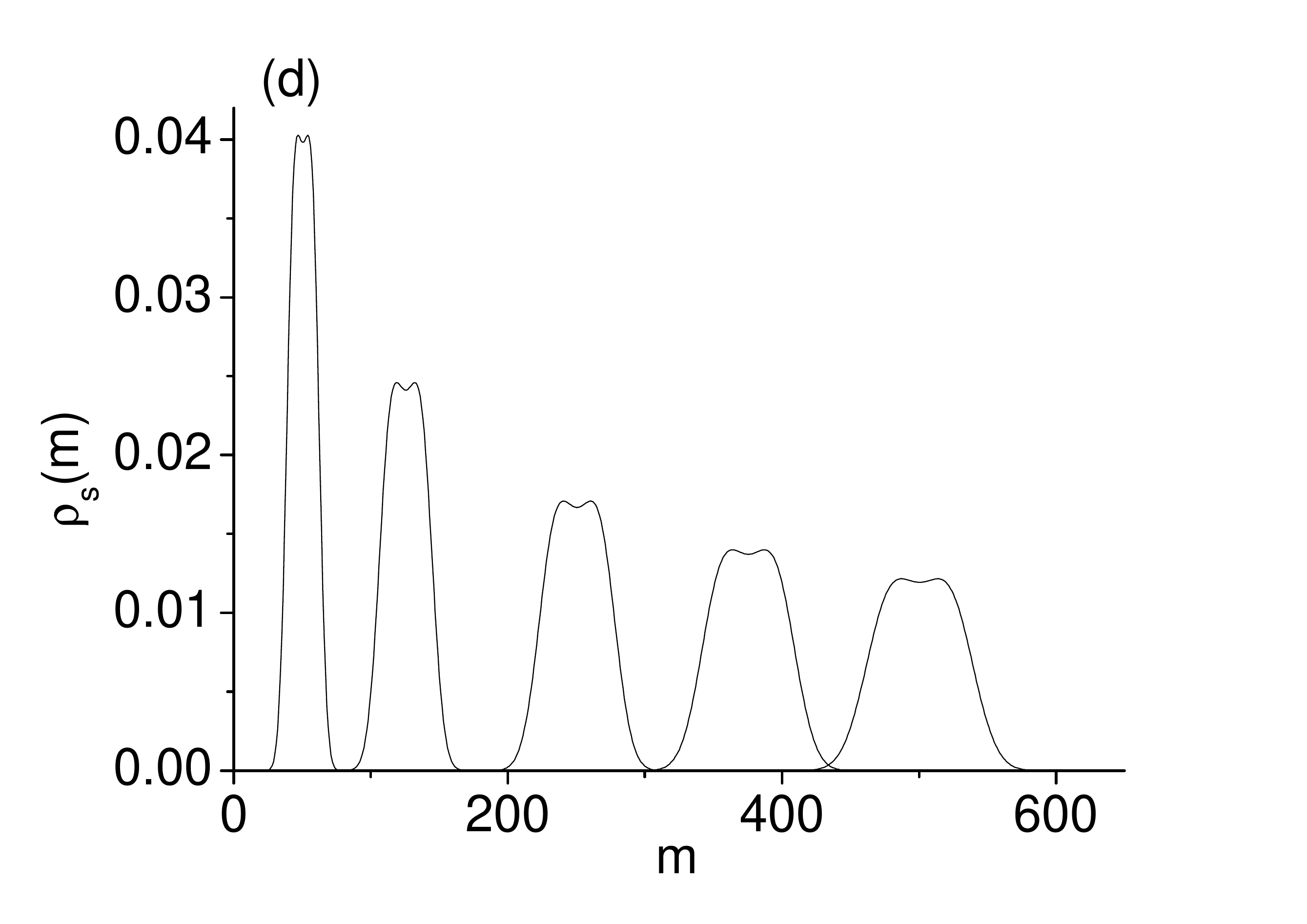}
   \caption{(color online) Comparison between numerical simulations (thick lines) and the
  approximate equilibrium distribution Eq.(\ref{prob_tot}) (thin lines). Panel
  (a), for $N_p=10$ also shows the result for a fully connected network with
  the rescaling Eq.(\ref{eq:rescaling}) corresponding to $f=99/2$ (red dashed
  curve). The dashed blue lines in panels (b) ($N_p=1$) and (c) ($N_p=0.1$)
  correspond to the approximation described in appendix \ref{appA}. The
  parameters are $T = 2  \times 10^4$, $p = 0.5$, $N = 100$. For the
  simulations $10^5$ realizations were performed. Panel (d) shows the
  distribution for different network sizes (from left to right $N=100$, $250$,
  $500$, $750$ and  $1000$) for $N_p=\sqrt{N}/4$, where the peak splits in
  two.}
  \label{fig3}
\end{figure}
\begin{equation}
\nu_1 = \frac{1+N_1}{1+N_0+N_1}.
\label{nu1}
\end{equation}
Therefore, the probability that $m$ nodes are in state 1 (the central node
plus $m-1$ peripheral nodes) becomes
\begin{equation}
p_1(m) = \binom{N}{m-1} \nu_1^{m-1} (1-\nu_1)^{N-m+1}.
\label{prob_r}
\end{equation}
Similarly, fixing the central node in the state 0, the asymptotic probability
that a peripheral node is in state 1 is
\begin{equation}
\nu_0 = \frac{N_1}{1+N_0+N_1},
\label{nu0}
\end{equation}
and the probability that $m$ nodes are in state 1 is
\begin{equation}
p_0(m) = \binom{N}{m} \nu_0^m (1-\nu_0)^{N-m}.
\label{prob_s}
\end{equation}
Adding these results we obtain the approximate expression
\begin{equation}
\rho_S(m) \approx \frac{N_1}{N_0+N_1} p_1(m) + \frac{N_0}{N_0+N_1} p_0(m),
\label{prob_tot}
\end{equation}
where we have introduced the weights $N_1/(N_0+N_1)$ and $N_0/(N_0+N_1)$ of the
central node to be in the state 1 or 0, respectively.

In this paper we will restrict our simulations to symmetric perturbations and
define, for simplicity, 
\begin{equation}
N_p \equiv N_0 = N_1.
\label{np}
\end{equation}

Fig.\ref{fig3} shows a comparison between simulations and the approximate
formula (\ref{prob_tot}). The two peaks are clearly related to the two states of
the central node and are reasonably well described by the approximation. The
region between the peaks is not well represented, since it has important
contributions from flips of the central node that have been discarded. The
dashed blue line shows the result of a better, although ad hoc, approximation
described in the appendix that fits the entire curve with very good precision.
Fig.\ref{fig3}(a) also shows the approximation (\ref{eq:rescaling}) obtained via
rescaling of the expression for fully connected networks, which works well for
$N_p >> 1$.

The approximate solutions can also be used to estimate the point where the
Gaussian-like distribution breaks in two peaks. For large $N$ the contributions
$p_0$ and $p_1$ for $\rho_S$ become Gaussians centered at $N \nu_0$ and $N
\nu_1$ with variance $\sigma^2 = N N_p (1+N_p)/(1+2N_p)^2$. The two-peak
structure appears when the distance between the two centers is of the
order of the standard deviation. This gives $N_p \sim \sqrt{N}$ and numerical
calculations indicate that $N_c \approx \sqrt{N}/4$. Fig.\ref{fig3}(d) shows the
equilibrium distribution for several values of $N$ and $N_p=\sqrt{N}/4$. The
transition from  unimodal to bimodal distribution marks the regime where the
influence of the central node competes with the external perturbation, modifying
the equilibrium distribution substantially with respect to the fully connected
dynamics. The two peaks move apart slowly as the external perturbation is
decreased and are clearly separated only when $N_p \sim 1$, independently of
the network size $N$. 

Although the equilibrium distribution of states of the star network changes
smoothly as the perturbation is decreased, the transition in behavior is rather
different from what is observed in the fully connected network: for $N_p >>
\sqrt{N}$ the state is disordered, with approximately half the nodes in state 1
and half the nodes in state 0. The standard deviation is $\sigma=\sqrt{N}/2$ so
that $\sigma/N=1/2\sqrt{N}$. For $N_p = \sqrt{N}/4$, when the two peak structure
appears, the standard deviation increases by a factor of 4 to
$\sigma/N=2/\sqrt{N}$. As $N_p$ decreases below 1 and the two peaks get
significantly apart, the network is most likely to be found with either a fraction
$\nu_1 = (1+N_p)/(1+2N_p)$  or $\nu_0 = N_p/(1+2N_p)$ in
state 1, executing fast collective transitions between the two states (see next
subsection). This is in contrast with the behavior exhibited by the fully
connected network, which have either most nodes 1 or most nodes 0 
staying in each of these states for long periods of time before
moving to the other.

% % % % % % % % % % % % % % % % % % % % % % % % % % % % % % % % % % %
\subsection{Dynamics and Magnetization}

In analogy with the Ising model we define the average magnetization per node as 
\begin{equation}
M = \frac{2 n_1}{N} - 1
\label{mag}
\end{equation}
where $n_1$ is the number of nodes in state 1, so that $-1 \leq M \leq +1$. In
order to study the dynamics of $M$ we run a single simulation for each network
and  plot $M$ as a function of the time.

\begin{figure}
   \includegraphics[clip=true,width=8.1cm]{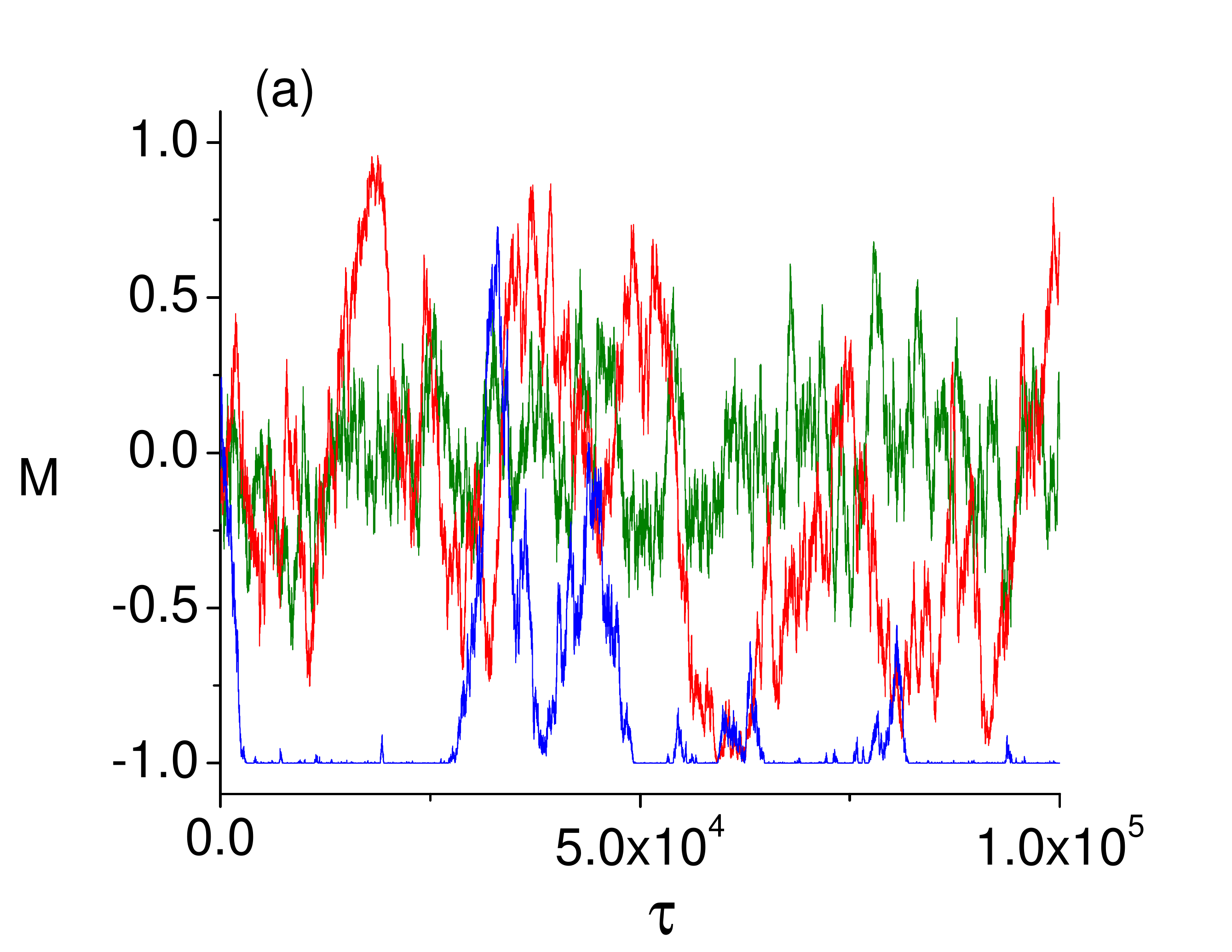}
   \includegraphics[clip=true,width=8.1cm]{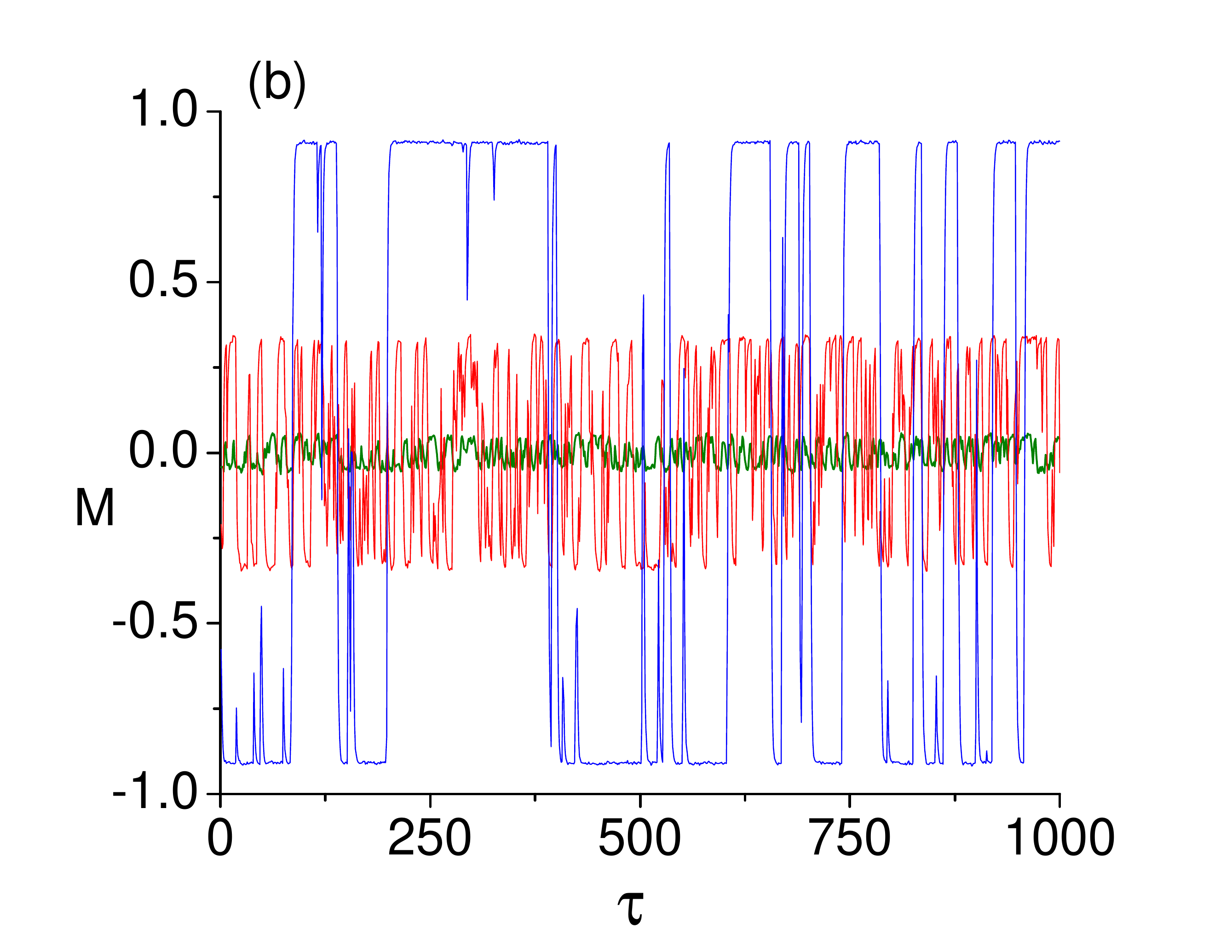}
   \caption{(color online) Magnetization for a single simulation for 
   $N_p=10$ (green), $N_p=1$ (red) and $N_p=0.05$ (blue).
   (a) fully  connected network with $N=2\times 10^4$ nodes;  (b) 
   star network with $N=2\times 10^4$ peripheral nodes. Time is measured in 
   units of network size.}
  \label{fig4}
\end{figure}

Fig. \ref{fig4} shows the results for $N_p =10$, $1$ and $0.05$ (see also Fig.
\ref{fig2}). In these plots one unit of time $\tau$ is a {\it Monte Carlo}
step, corresponding to $N$ steps $t$ of the dynamics, so that all nodes are
updated, on average, at each unit of $\tau$. For the fully connected network
with $N=20000$, Fig.\ref{fig4}(a), $M$ fluctuates around zero for $N_p = 10$
(green line). The fluctuations increase as the critical value is approached and
for $N_p = 1$ (red line) they take the entire range of $M$. For $N_p=0.05$ (blue
line) the system stays a substantial amount of time {\it magnetized} at $M=+1$
or $M=-1$, alternating from one extreme to the other. The lower the values of
$N_p$ the longer the times the system stays in each state for a fixed value of
$N$, and similarly for increasing $N$ for fixed $N_p$.

For the star network, Fig.\ref{fig4}(b), the results show two distinct features.
First, the amplitude of the oscillations increases smoothly as $N_p$ decreases,
reflecting the position of the two peaks of the equilibrium distribution. For
$N_p=1$, for instance, $M$ oscillates in the interval around $\pm 0.3$. Second,
the oscillations are much faster, on the scale of tens of time steps for
$N_p=0.05$, as compared to the thousands of time steps of the fully connected
network. These oscillations are clearly driven by flips of the central node,
which pulls the majority of the peripheral nodes with it. 

\begin{figure}
   \includegraphics[clip=true,width=12cm]{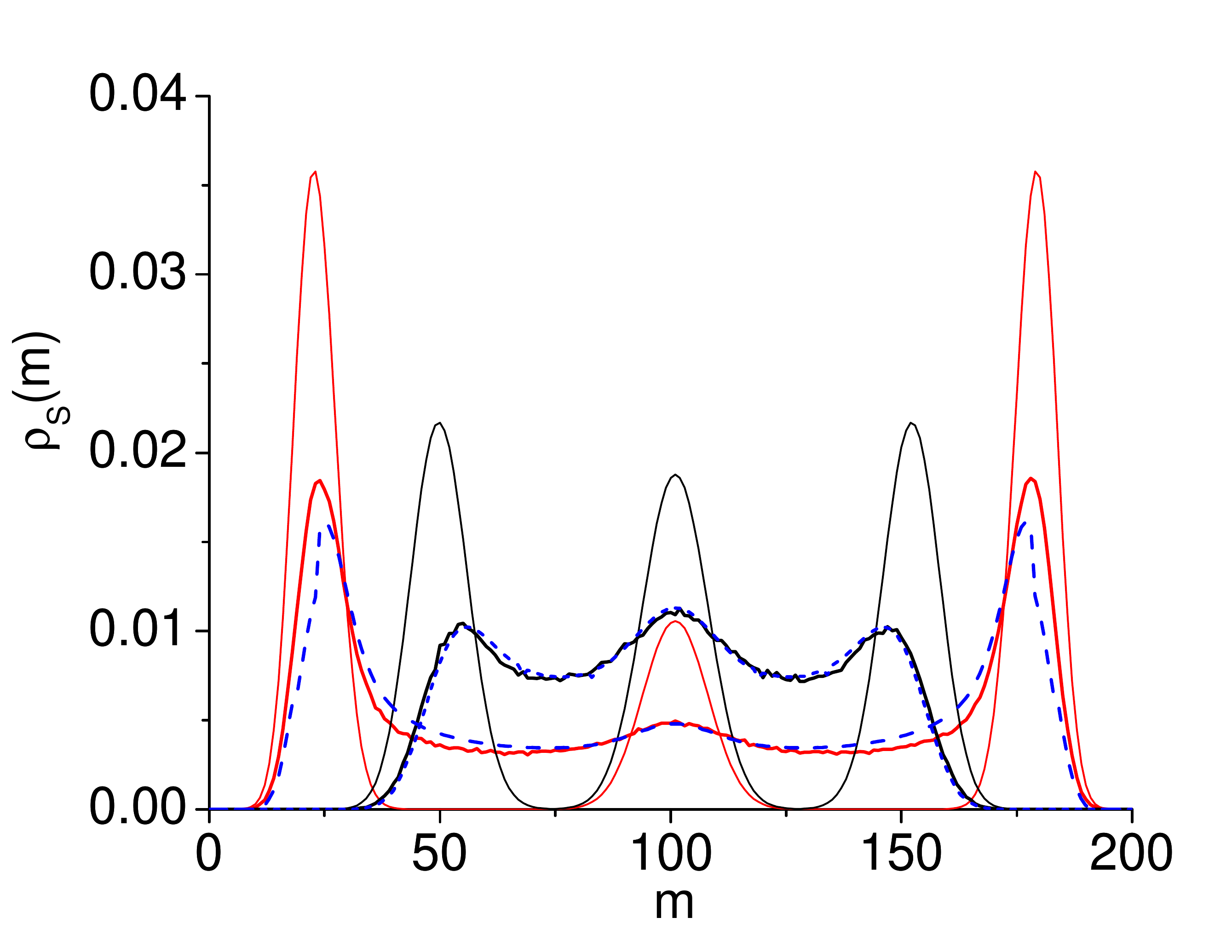}
   \caption{(color online) Equilibrium probability distribution $\rho_S(m)$ for
   a star network with $N_c=2$ center nodes, $N=200$ peripheral nodes and
   $N_0=N_1=N_p$ for $N_p=1$ (black) and $N_p=0.3$ (red). Thick curves show the
   result of simulations and thin curves the approximation given by
   Eq.(\ref{prob_totg}). The dashed blue lines correspond to the approximation
   described in appendix \ref{appA}.}
  \label{fig5}
\end{figure}

The large difference in the time scales displayed in Fig. \ref{fig4} can be
understood from the network topologies. For fully connected networks the time
scale measured in number of time steps $t$ is well known, given by
$t=N(N+2N_p-1)/2N_p$ (see \cite{chinellato2015}, for instance). For small $N_p$
we find $\tau \equiv t/N \simeq N/2 N_p$. For the star network, on the other
hand, the state of the peripheral nodes is controlled by the central node. If
the central node is in state $0$, most of the peripheral nodes will be in state
$0$ as well if $N_p$ is small. The probability that the central node flips from
$0$ to $1$ can be estimated as the probability that it copies a frozen node in
state $1$: $N_p/(N-1+2N_p) \simeq N_p/N$. The average time for this to happen
is $t = N/N_p$ or $\tau = t/N = 1/N_p$. The two time scales differ by a factor
$N/2$, which is consistent with the results shown by Fig. \ref{fig4}.

% % % % % % % % % % % % % % % % % % % % % % % % % % % % % % % % % % %
\subsection{Generalizations}

Star networks where the center is composed not by a single node, but by a group
of totally connected nodes can also be studied within this approximation. If the
center has $N_c$ nodes a stationary solution can be constructed by freezing the
state of the center into $m$ ones and $N_c-m$ zeros and assigning a weight to
this state according to the fully connected distribution $\rho_{FC}(m)$, given
by Eq.(\ref{probn}). Equation (\ref{prob_tot}) readily generalizes to
\begin{equation}
\rho(m) \approx \sum_{k=0}^{N_c} \rho_{FC}(k) \binom{N}{m-k}
\nu_k^{m-k} (1-\nu_k)^{N-m+k},
\label{prob_totg}
\end{equation}
where
\begin{equation}
\nu_k = \frac{N_1+k}{N_c+N_0+N_1}
\label{nuk}
\end{equation}
and $\rho(k)_{FC}$ is given by equation (\ref{probn}) with $N$ replaced by $N_c$.
Fig. \ref{fig5} shows an example with $N_c=2$ where a three peak structure is
clearly visible close to the phase transition $N_0=N_1=1$. The approximation
(\ref{prob_totg}) captures well the position of the peaks, but overshoots their
height to compensate for the lost interference between the peaks.

As a second application we consider the joint effect of two hubs in a complex
network. If we approximate the hubs as independent star networks with a
single central node, the probability of finding $m$ nodes in the state 1 is
simply given by  
\begin{equation}
\rho(m) = \sum_{j=0}^m \rho_{S, \,{\cal N}_1}(j)  \rho_{S, \,{\cal N}_2}(m-j),
\label{prob12}
\end{equation}
where we have indicated explicitly the number of nodes of each star in the
distribution. For small values of $N_p$, the separate distributions will have
two peaks, centered at, say, $m_1$ and ${\cal N}_1-m_1$; $m_2$ and ${\cal
N}_2-m_2$ respectively. The joint distribution given by Eq.(\ref{prob12}) will
display four peaks at $m_1+m_2$, ${\cal N}_1-m_1+m_2$, ${\cal N}_2-m_2+m_1$ and
${\cal N}_1+{\cal N}_2-m_1-m_2$. If the hubs are not independent, but coupled by
only a few links, we expect this peak structure to persist.

\begin{figure}
   \includegraphics[clip=true,width=8.1cm]{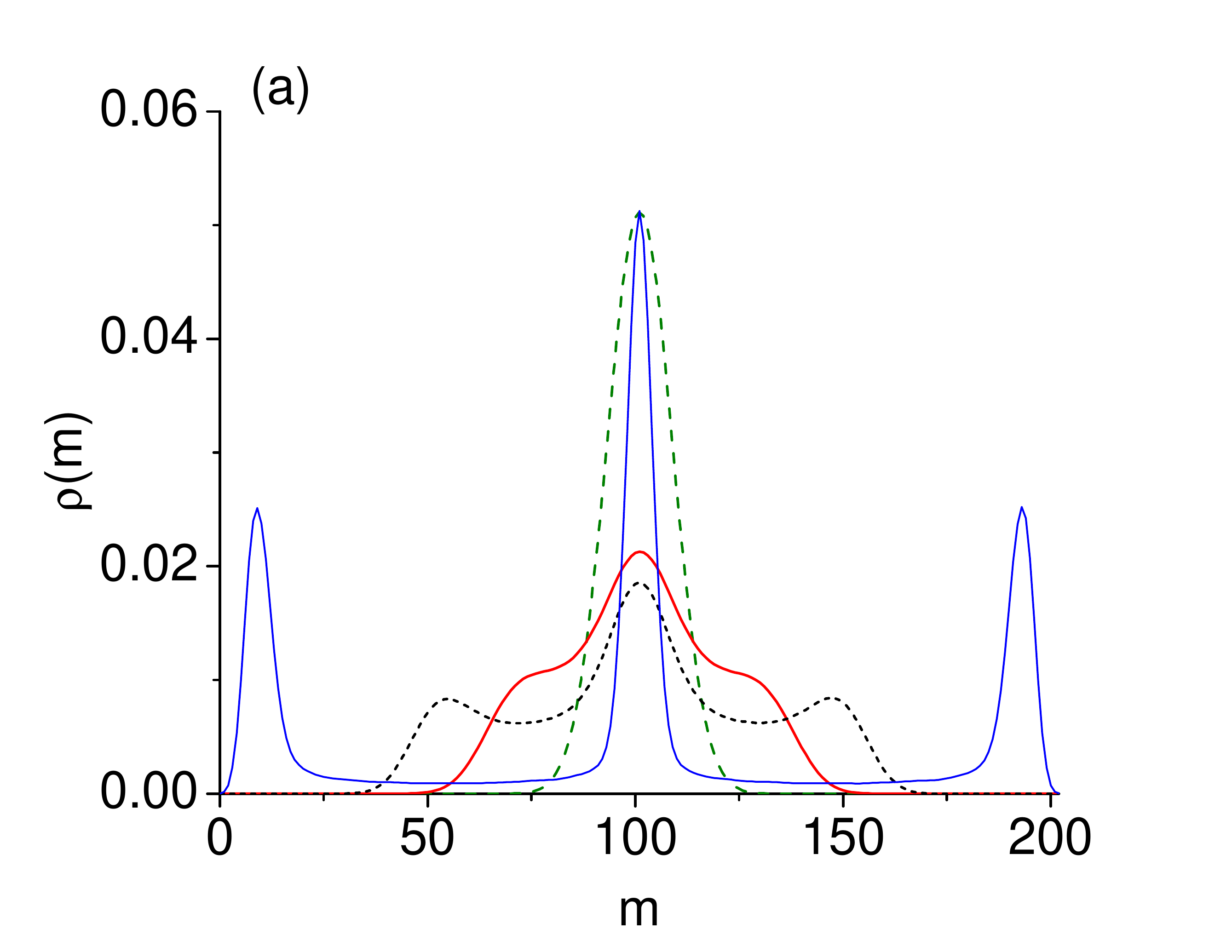}
   \includegraphics[clip=true,width=8.1cm]{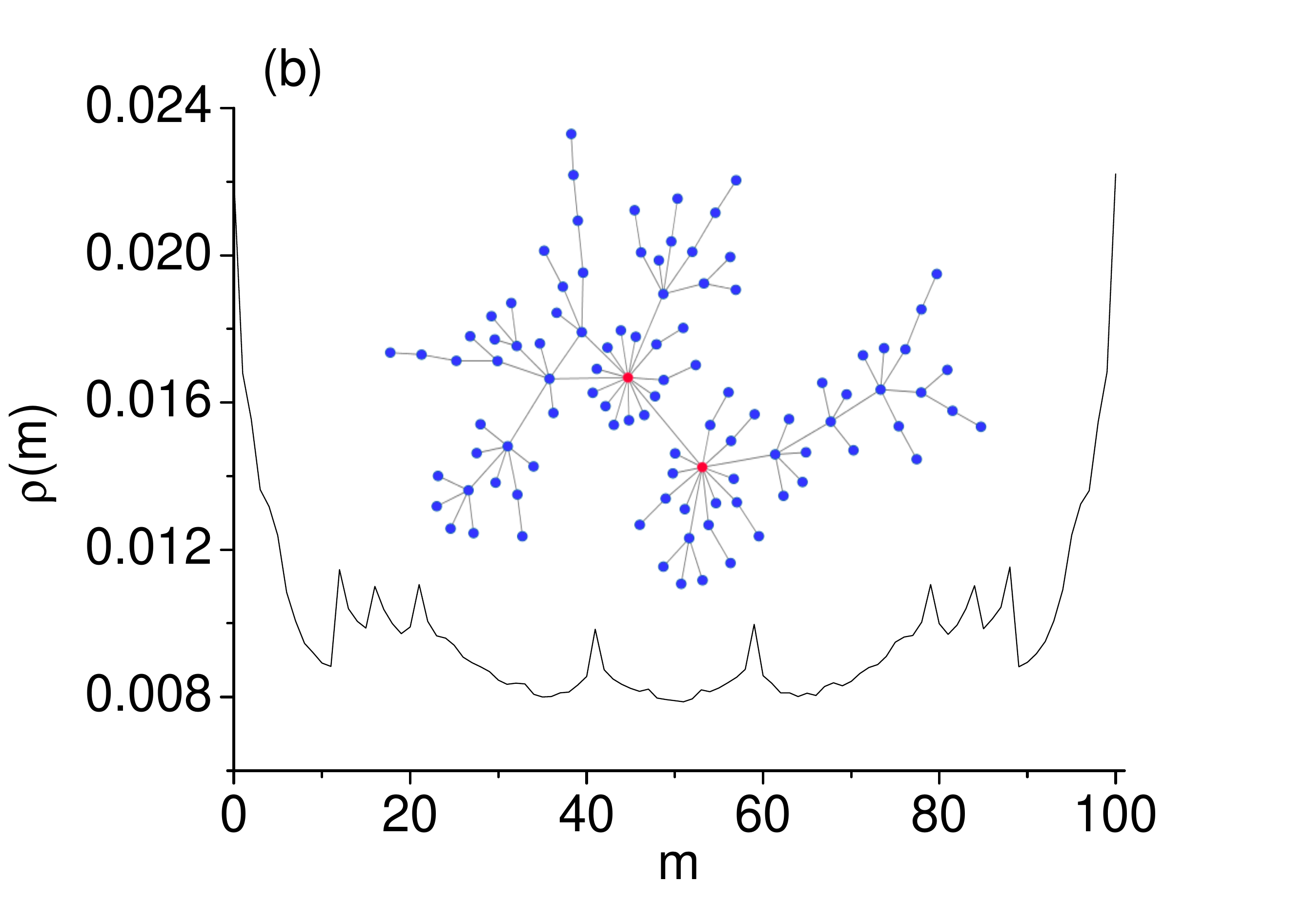}
   \caption{(color online) (a) Equilibrium distribution for a pair of
  independent star networks with ${\cal N}_1=101$ and ${\cal N}_2=11$ nodes
  for $N_p=10$ (green dashed), $N_p=1$ (thick red), $N_p=0.5$ (black dotted) and
   $N_p=0.05$ (blue). 
  (b) Distribution for a scale-free network  with $100$ nodes and
  $N_p=0.01$. The two largest hubs (shown in red in the inset)
  have 16 and 13  peripheral nodes and are connected by their centers.}
  \label{fig6}
\end{figure}

Figure \ref{fig6}(a) shows the stationary distribution for two independent
star networks of sizes ${\cal N}_1=101$ and ${\cal N}_2=11$. The 
splitting of the Gaussian-like peak in two occurs at $N_0=N_1\approx 2.6$ and 
$N_0=N_1\approx 0.8$, respectively. However, because the separation between the
two peaks of the smaller star is small, its effect is felt only at much smaller
values of the perturbation, when the two peaks of the large hub approach the
borders and the distribution becomes thin. Fig. \ref{fig6}(b) shows the
equilibrium distribution for a more complex network with 100 nodes constructed
with preferential attachment. The network has two main hubs (shown in red in the
inset) with 16 and 13 peripheral nodes, respectively. The peaks in the
distribution are signatures of the hubs. For scale-free networks with more
cycles (not shown) the presence of the peaks is much less conspicuous and the
distribution becomes again similar to the fully connected case with rescaled
perturbations.

%%%%%%%%%%%%%%%%%%%%%%%%%%%%%%%%%%%%%%%%%%%%%%%%%%%%%%%%%
\section{Conclusions}\label{conclusions}

The voter model with opinion makers is one of the simplest dynamical systems
that can be represented on a network. It models an election between candidates
where the voters are influenced by their social contacts and by external
factors such as journalists and politicians. If the number of opinion makers is
zero the population is certain to reach a consensus toward one of the candidates
independently of the structure of the network. The power of the opinion makers,
however, depends strongly on the average degree of the network. For a completely
connected network the transition between nearly consensus (ordered state) and a
tie (disordered state) takes place exactly at $N_0=N_1=1$, independently of the
population size $N$. For networks with average degree $k_{av}$ the effect of the
fixed nodes is amplified by a factor $f=(N-1)/k_{av}$, which can be very large
for natural populations. Much above or much below the transition from disordered
to ordered states the influence of the network structure is negligible and only
shows up in the rescaling of the opinion makers influence, that is large when
the network is weakly connected. This is in contrast with processes describing
the spreading of epidemics or synchronization of oscillators, where the topology
plays a crucial role
\cite{barrat08,ves01,pecora02,motter03,pac04,laguna,guime,aguiar05,rodrigues13,rodrigues14}. 
Close to the critical point, however, the network structure can leave
signatures in the probability distribution $\rho(m)$.

For the particular case of star networks with a single central node, the
Gaussian-like distribution displayed by $\rho_{S}(m)$ for large values of $N_0$
and $N_1$ splits into two peaks centered at $N(N_1+1)/(N_1+N_0+1)$ and $N
N_1/(N_1+N_0+1)$ reflecting the state of the central node being 1 or 0. The
central node controls the entire system and the distribution behaves
approximately as a single giant node with two collective states only. For
$N_0=N_1=1$ the peaks are centered at $2N/3$ and $N/3$ respectively, which is
rather different from the distribution of fully connected networks where
$\rho_{FC}(m)=1/(N+1)$ is constant. In the former case the election will be won
by one of the candidates with approximately $67\%$ of the votes, whereas in the
latter, the winner can have any number of votes with equal probability. 
For small values of $N_0$ and $N_1$ both star and fully connected networks are
likely to be found in ordered states, where most nodes are in state 0 or in
state 1. These states, however, are not stable and network oscillates between
the two possibilities. We found that the average frequency of these
oscillations are much higher for star networks than for fully connected ones. 

When a few weakly connected hubs are present, the effects of central nodes are
still visible, as shown by Fig.\ref{fig6}(b). However, when the
system is controlled by multiple hubs, as in a general scale-free network, the
collective behavior becomes again similar to that predicted by the mean field
approximation and the control by `local leaders'  becomes much less relevant. In
these cases Eqs.(\ref{probn}) and (\ref{eq:rescaling}) provide good
approximations for the equilibrium probability.

As a final remark we note that fully connected and stars with arbitrary number
of central nodes seem to be the only network topologies where a simple treatment
via macroscopic master equations similar to (\ref{eqr})-(\ref{eqs}) is
possible. Even the highly symmetric ring network (1-D lattice with periodic
boundary conditions) does not behave as if all nodes were identical, since 
different configuration having the same number of nodes at the state 1 give rise
to different macroscopic states.\\

M.A.M.A. and D.M.S. acknowledge financial support from CNPq and FAPESP. C.A.M.
was supported by CAPES.

\newpage
%%%%%%%%%%%%%%%%%%%%%%%%%%%%%%%%%%%%%%%%%%%%%%%%%%%%%%%%%%%
%%%%%%%%%%%%%%%%%%%%%%%%%%%%%%%%%%%%%%%%%%%%%%%%%%%%%%%%%%%
\begin{appendix}

%%%%%%%%%%%%%%%%%%%%%%%%%%%%%%%%%%%%%%%%%%%%%%%%%%%%%%%%%%%
\section{An ad hoc approximation for the equilibrium distribution}
\label{appA}

The approximation (\ref{prob_tot}) completely discards the fact that the state
of the central node fluctuates and fails to describe the region between the two
peaks. Here we derive a better approximation using phenomenological ideas. We
first define
\begin{equation}
\nu(x) = \frac{x+N_1}{1+N_0+N_1}.
\label{nux}
\end{equation}
as the equivalent of (\ref{nu1}) and (\ref{nu0}) for the case where the state
of the central node is in the average state $x$ with $0 \leq x \leq 1$.
Accordingly, we define
\begin{equation}
p(x,m) = \frac{\Gamma(N+1)}{\Gamma(N-m+x+1) \Gamma(m-x+1)} \nu(x)^{m-x}
(1-\nu(x))^{N-m+x}
\label{probx}
\end{equation}
as the probability of finding $m$ nodes in state 1, including the central the
peripheral nodes (see eqs. (\ref{prob_s}) and (\ref{prob_r})). If $c(x)$ is the
probability distribution that the central node is in state $x$, then
\begin{equation}
\rho(m) = \int_0^1 dx \, c(x,N_0,N_1) p(x,m).
\label{probx0}
\end{equation}
For 
\begin{equation}
c(x,N_0,N_1) = \frac{N_0}{N_0+N_1}\delta(x) + \frac{N_1}{N_0+N_1}\delta(x-1)
\label{cx0}
\end{equation}
we recover the approximation (\ref{prob_tot}). 

In order to obtain better results we need to consider smoother distributions and
the natural functional dependence for $c(x)$ is the continuous version of
$\rho_{FC}$, the Beta distribution eq.(\ref{probbetax}):
\begin{equation}
c(x,N_0,N_1) =
\frac{\Gamma(n_0+n_1)}{\Gamma(n_0) \Gamma(n_1)} x^{n_0-1}
(1-x)^{n_1-1}.
\label{cbeta}
\end{equation}
Here $n_0(N_0,N_1)$ and $n_1(N_0,N_1)$ measure the joint effect of the external
perturbations, $N_0$ and $N_1$, and of the $N-1$ peripheral nodes on the central
node. Because the approximation with the delta functions (\ref{cx0}) already
gives a good description of the exact distribution, $n_0$ and $n_1$ should be
significant only close to the phase transition. The choice
\begin{equation}
n_0(N_0,N_1) = N_0 e^{-(N_0+N_1)/2} \qquad n_1(N_0,N_1) = N_1 e^{-(N_0+N_1)/2} 
\end{equation}
turns out to work well for all the cases tested.

For the case of two central nodes (see Fig.\ref{fig5}) a similar procedure can
be divised. We set
\begin{equation}
\nu(x,y) = \frac{x+y+N_1}{2+N_0+N_1}
\label{nux2c}
\end{equation}
with $x$ and $y$ representing the states of the two central nodes. The
probability that $m$ nodes are in state 1 becomes
\begin{equation}
p(x,y,m) = \frac{\Gamma(N+1)}{\Gamma(N-m+x+y+1) \Gamma(m-x-y+1)}
\nu(x,y)^{m-x-y} (1-\nu(x,y))^{N-m+x+y}
\label{probx2c}
\end{equation}
so that
\begin{equation}
\rho(m) = \int_0^1 dx \, \int_0^1 dy \, c(x,y,N_0,N_1) p(x,y,m).
\label{probxy0}
\end{equation}
The probability distribution that the two central nodes are in states $x$ and
$y$ must reproduce the coefficients $\rho_{FC}(k)$ in Eq.(\ref{prob_totg}).
Using the analogy between Eqs. (\ref{prob_tot}) and (\ref{cx0}) it can be
checked that the appropriate function is
\begin{equation}
c_2(x,y,N_0,N_1) = c(x,N_0,N_1) \,  c(y,N_0+1-x,N_1+x).
\label{c2beta}
\end{equation}
Indeed, using the approximation (\ref{cx0}) for $c(x,N_0,N_1)$ we see that
\begin{equation}
\begin{array}{ll}
c_2(x,y,n_0,n_1) & \approx \frac{1}{(N_0+N_1)(1+N_0+N_1)} \times
\left\{ N_0(N_0+1) \delta(x) \delta(y) \right.\\ \\
& \left. + N_0 N_1 [\delta(x) \delta(1-y) +
\delta(1-x) \delta(y)] + N_1(N_1+1) \delta(1-x) \delta(1-y) \right\}
\end{array}
\end{equation}
whose coefficients correspond to $\rho_{FC}(k)$. We remark that the integrals
(\ref{probx0}) and (\ref{probxy0}) might be difficult to evaluate numerically
for very small values of $N_0$ and $N_1$, since the Beta distribution becomes
very large close to $x=0$ and $x=1$. In this limit, however, the distribution is
peaked close to $m=0$ and $m=N$ and the approximation provided by the fully
connected distribution should work well. 

\end{appendix}

%%%%%%%%%%%%%%%%%%%%%%%%%%%%%%%%%%%%%%%%%%%%%%%%%%%%%%%%%

\end{document}